\documentclass[sigconf]{acmart}

\AtBeginDocument{%
  \providecommand\BibTeX{{%
    \normalfont B\kern-0.5em{\scshape i\kern-0.25em b}\kern-0.8em\TeX}}}

\copyrightyear{2020} 
\acmYear{2020} 
\setcopyright{acmcopyright}\acmConference[CIKM '20]{Proceedings of the 29th ACM International Conference on Information and Knowledge Management}{October 19--23, 2020}{Virtual Event, Ireland}
\acmBooktitle{Proceedings of the 29th ACM International Conference on Information and Knowledge Management (CIKM '20), October 19--23, 2020, Virtual Event, Ireland}

\usepackage{graphicx}
\usepackage{amsmath}
\usepackage{bm}
\usepackage{amsfonts}
\usepackage{amsthm}
\usepackage{multirow}
\usepackage{colortbl}
\usepackage{enumitem}
\usepackage{subcaption}
\usepackage{array}
\usepackage{caption}
\usepackage[autolanguage]{numprint}

\usepackage{xcolor}

\definecolor{bettergreen}{RGB}{80, 191, 82}

\newcommand{\reducedstrut}{\vrule width 0pt height .9\ht\strutbox depth .9\dp\strutbox\relax}
\newcommand{\cb}[2]{%
  \begingroup
  \setlength{\fboxsep}{0pt}%
  \colorbox{bettergreen!#1}{\reducedstrut#2\/}%
  \endgroup
}

\begin{document}
\title{Fine-Grained Relevance Annotations for Multi-Task Document~Ranking~and~Question~Answering}

%
%
\author{Sebastian Hofst{\"a}tter, Markus Zlabinger, Mete Sertkan, Michael Schr{\"o}der and Allan Hanbury}
\email{first.last@tuwien.ac.at}
\affiliation{\institution{TU Wien, Vienna, Austria}}

\begin{abstract}
There are many existing retrieval and question answering datasets. However, most of them either focus on ranked list evaluation or single-candidate question answering. This divide makes it challenging to properly evaluate approaches concerned with ranking documents and providing snippets or answers for a given query. In this work, we present FiRA: a novel dataset of Fine-Grained Relevance Annotations. We extend the ranked retrieval annotations of the Deep Learning track of TREC 2019 with passage and word level graded relevance annotations for all relevant documents. We use our newly created data to study the distribution of relevance in long documents, as well as the attention of annotators to specific positions of the text. As an example, we evaluate the recently introduced TKL document ranking model. We find that although TKL exhibits state-of-the-art retrieval results for long documents, it misses many relevant passages. 
\end{abstract}

\maketitle
\section{Introduction}





\begin{figure}
    \centering
    \vspace{0.4cm}
    \includegraphics[width=0.475\textwidth,trim=0 2.1cm 0 0, clip ]{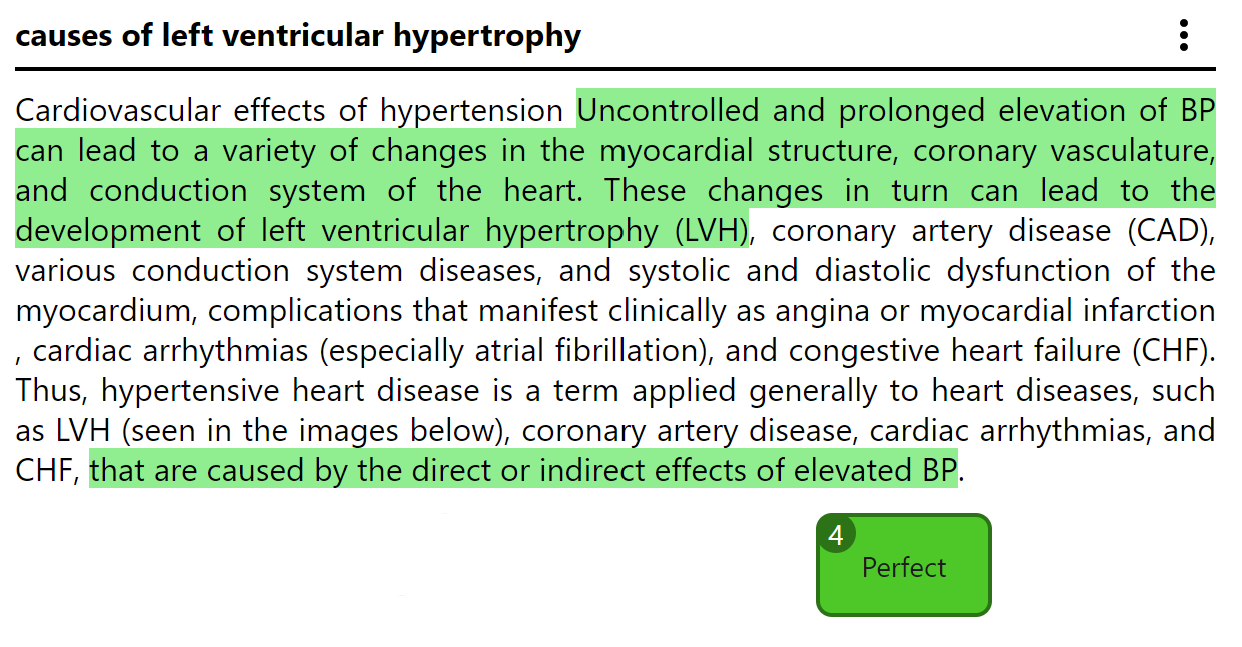}
    \vspace{-0.6cm}
    \caption{This screenshot of our FiRA tool shows the task of selecting fine-grained relevance spans (or answers) within a relevant document snippet.}
    \label{fig:fira-screenshot}
    \vspace{-0.5cm}
\end{figure}

Currently, there is a substantial disconnect between the way neural retrieval models operate on full text and the relevance labels that guide learning, evaluation and analysis. Those labels, such as in the TREC Deep Learning Track 2019 (TREC-DL) \cite{trec2019overview}, cover a long document, with thousands of tokens, as a whole. For traditional models, which collapse document statistics to a bag of words, this label granularity is sufficient. However, recently introduced neural ranking models operate on a more fine-grained level. They hierarchically build up their score from interactions of all words to a single score for the full document \cite{Hofstaetter2020_sigir,zheng2019rltm,yilmaz2019}. Therefore, the task of neural retrieval models shifts towards a combination of document-level relevance and word-level question answering or information extraction. The models are simultaneously tasked to detect relevant spans of information and, based on these, make a scoring decision that can be used to rank many candidate documents. However, a considerable bottleneck in their development and evaluation is the lack of datasets covering both sub-tasks equally well. Current datasets either focus on retrieval results with dense judgements across ranked lists \cite{trec2019overview,hashemi2020antique}, or question answer selections in single candidate texts that can retroactively be converted to retrieval collections, but lead to incomplete retrieval judgements \cite{rajpurkar2016squad}. 

In this work, we present FiRA: a new dataset of \textbf{Fi}ne-grained \textbf{R}elevance \textbf{A}nnotations with word-level relevance selections and passage-level graded relevance labels for all relevant documents of the TREC 2019 Deep Learning Document dataset. We split the documents into snippets and displayed query \& document snippet pairs to annotators. In Figure \ref{fig:fira-screenshot}, we show an example of an annotated document-snippet. We ensure a high quality by employing at least 3-way majority voting for every candidate and continuous monitoring of quality parameters during our annotation campaign, such as the time spent per annotation. Furthermore, by requiring annotators to select relevant text spans, we reduce the probability of false-positive retrieval relevance labels.
We conducted our annotation campaign with 87 computer science students using a custom annotation tool, which provides a seamless workflow between selecting relevance labels and relevant text spans.

The FiRA dataset contains \numprint{24199} query \& document-snippet pairs of all \numprint{1990} relevant documents for the \numprint{43} queries of TREC-DL. We chose to extend the TREC-DL document ranking test collection and use the ranked retrieval data as a starting point, because 1) it allows us to narrow our annotation task, as fine-grained annotations are very time-consuming and 2) we are able to compare our non-expert annotators to TREC's expert annotators. FiRA is not specific to any model, but incorporates the full pool of TREC annotations. 

Our novel FiRA dataset enables a variety of scenarios:
\begin{itemize}
    \item \textbf{Training:} FiRA can be used for standalone and multi-task retrieval \& QA fine-tuning. The density and coverage over the query-document pairs allow for granular fine-tuning using word-level loss functions.
    \item \textbf{Evaluation:} FiRA augments the TREC-DL evaluation, with more fine-grained relevance labels and the ability to evaluate the answer passages extracted from the full documents. Combined with the labeled non-relevant documents of TREC-DL, FiRA allows to evaluate rankings and answer selections.
    \item \textbf{Analysis:} Neural document ranking models that output relevance scores for a full document can be thoroughly analyzed. Especially, their partial results and attention regions can be quantitatively inspected with FiRA.
\end{itemize}

The human subjectivity of the relevance task is a recurring topic in IR research \cite{borlund2003concept}. When using human-annotated datasets, it is important to know about the noise and uncertainty that different human relevance judgements bring. For this, we employ 3-way majority voting throughout our data collection. Additionally, we deliver our FiRA dataset with a quantitative analysis of the question: 
\begin{itemize}
    \item[\textbf{RQ1}] How much subjectivity do our fine-grained relevance annotations exhibit?
\end{itemize}
To that end, we selected 10 pairs to be annotated by all our annotators, so that we can study their subjectivity with a dense distribution that reduces outliers. We found, that on a 2-level-graded relevance scale our annotators agree strongly, whereas on the 4-level-graded relevance and the text selections the subjectivity increases.

We utilize our new dataset and present a thorough study of the positional bias in both relevance across a document and inside a single snippet.
By providing annotators snippets in random order, so that there is no bias of attention towards the beginning, we aim to answer our research question:
\begin{itemize}
    \item[\textbf{RQ2}] Does there exist a position-based relevance bias in long documents?
\end{itemize}
Our FiRA annotations show that there is indeed an increased likelihood of encountering relevant answers at the beginning of documents, although we also observe that there is a considerable number of relevant snippets found in later parts of the documents.

Furthermore, we studied the behavior of our annotators concerning the selection of relevant words inside a single snippet. An analysis of the MSMARCO-QA \cite{msmarco16} dataset showed a strong bias towards the beginning of snippets. Therefore, using our FiRA annotations, we answer the question:
\begin{itemize}
    \item[\textbf{RQ3}] Are annotators biased towards the beginning of a snippet?
\end{itemize}
For this we created a control-group and a treatment-group, in which we rotated the first and second halves of the texts. Neither group shows a position bias towards the start. The rotation-group actually amplifies the importance of context around a relevant span selection, as split sentences were less likely to be selected. 

To showcase the usefulness of FiRA, we use the fine-grained judgements to analyze how well the recently introduced state-of-the-art TKL neural document re-ranking model \cite{Hofstaetter2020_sigir} reaches its potential by utilizing TKL's interpretability functions. The TKL model provides information about which three regions with 30 words of a document it used compute the score.
\begin{itemize}
    \item[\textbf{RQ4}] How many of FiRA's relevant spans does the TKL model use to compute the document score?
\end{itemize}
We find that the TKL model trained on TREC-DL data \cite{trec2019overview} has a disproportionate focus on the beginning of documents as a ranking signal. 

The observations made using FiRA can inspire the next generation of neural document ranking models. Our FiRA dataset, including raw annotations, documentation and all accompanying helper scripts, is freely available at: \\
\textit{\url{https://github.com/sebastian-hofstaetter/fira-trec-19-dataset}}

\section{Methodology}

We describe how we transformed the TREC-DL dataset to prepare it for fine-grained annotations; our annotation task description; followed by timeline analysis of the FiRA annotation exercise that we conducted among computer science students.

\subsection{Data Preparation}

Our dataset is based on the document ranking task of TREC's Deep Learning track 2019. This guarantees a broad and diverse pool of documents in our annotation campaign. TREC used the following graded relevance labels: \textit{Not Relevant} (0), \textit{Relevant} (1),
\textit{Highly Relevant} (2), and \textit{Perfect} (3). For the 43 queries of the DL task we gathered all documents marked as \textit{Highly Relevant} (1149 query-document pairs) and \textit{Perfect} (841 query-document pairs) by the TREC experts. We concatenated the title and body text, and did not visually indicate the title to the annotators. 

The documents are long with an average \numprint{3039} words and a median of \numprint{1171} words. We assumed that if we want exact and exhaustive word-level relevance annotations, we need to divide the documents into smaller pieces, as this has been shown to be an effective task simplification \cite{Finnerty2013}. We therefore split the documents into snippets. We aimed to create semantically coherent snippets of complete sentences of approximately 120--130 words,\footnote{We split by whitespace, other tokenization methods may result in more tokens.} as this is the maximum number of words that fit on a typical smartphone screen without the need to scroll. We used the following rules to create our snippets:

\begin{itemize}
    \item Split a document based on sentences using the BlingFire library.\footnote{https://github.com/microsoft/BlingFire}
    \item Add sentences to a snippet until the maximum length of 130 words is reached. If it overflows, we start with a new snippet starting with this sentence.
    \item If a sentence does not fit (e.g. if a document does not contain any punctuation), we split by whitespace, so we could concatenate tokens back together, to fit the desired snippet length.
\end{itemize}

We created a maximum of 30 document snippets per document, resulting in a cap of roughly 4,000 tokens, as this was the maximum capacity of our annotation resources. We opted to prioritize complete majority voting coverage over more depth. 

This procedure resulted in a total of \numprint{24199} document snippets, with an average of \numprint{12} snippets per query-document pair.

\begin{figure*}
    \centering
    \includegraphics[width=0.95\textwidth]{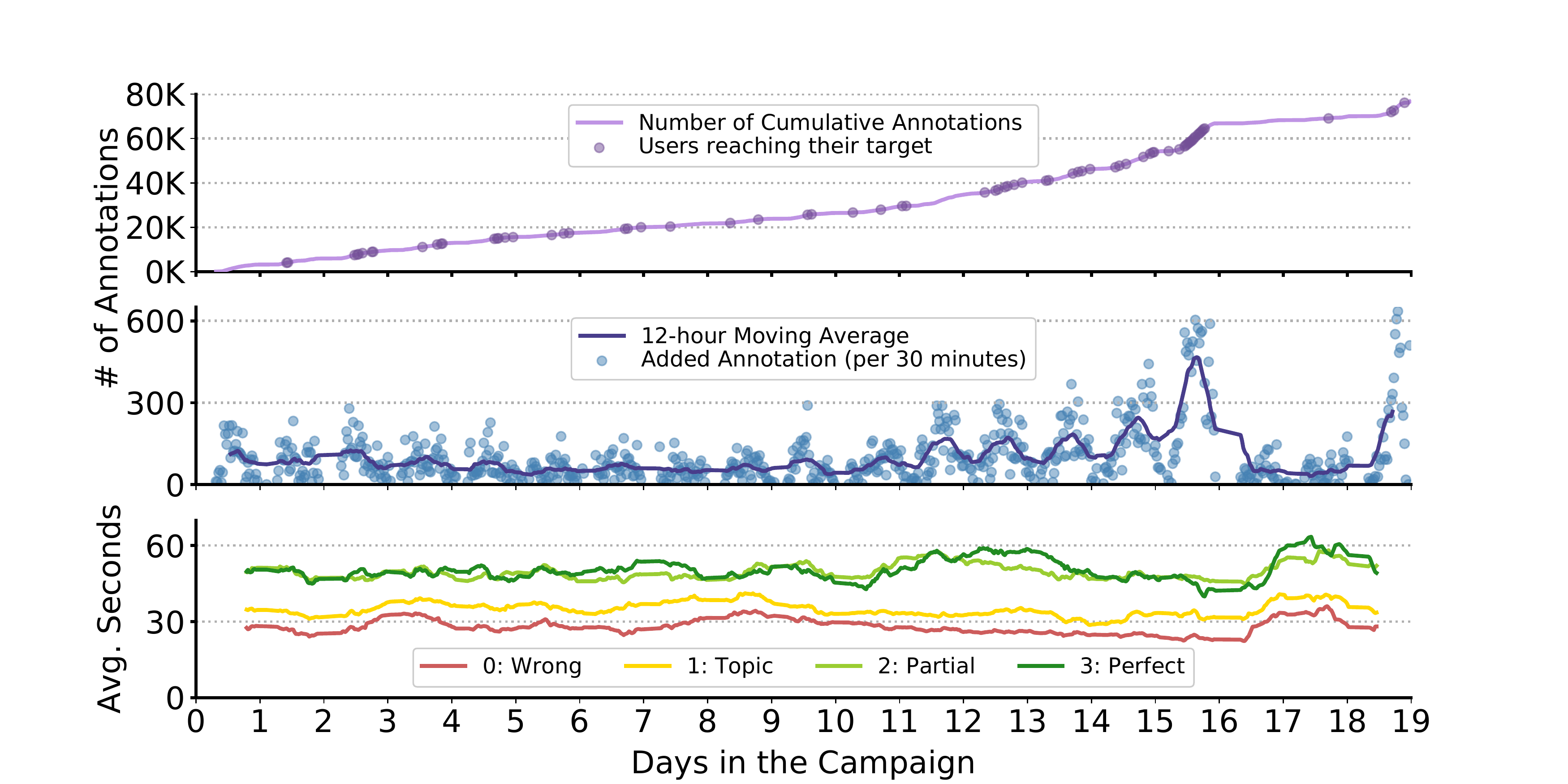}
    \caption{Monitoring of our annotation campaign. The topmost plot shows the total annotation count; the middle plot the added annotations in a 30 minute window; and the bottom plot shows the daily-average time spent per annotation class. Note: The first deadline was set to the end of day 16 and later extended by three days.}
    \label{fig:campaign-monitoring}
\end{figure*}

\subsection{FiRA Annotation Task Guidelines}

\begin{figure}
    \centering
    \includegraphics[width=0.45\textwidth]{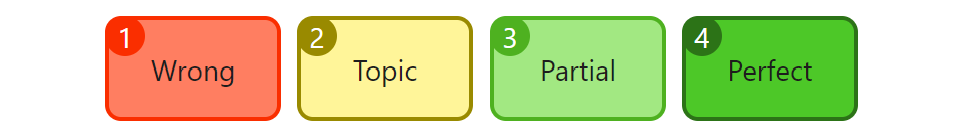}
    \caption{Screenshot of the label-selection UI of the FiRA annotation tool. The numbers indicate keyboard shortcuts.}
    \label{fig:labels}
    \vspace{-0.4cm}
\end{figure}

The guiding principle of this annotation campaign was to divide the task into the smallest possible pieces and allow for desktop as well as mobile annotations, to allow annotation to take place in the widest possible types of settings. Our aim was to create a simple and reduced user interface, which included condensing the naming schema of the relevance classes to fit on a single line on a smartphone. We kept the graded relevance from TREC, however we re-named the relevance labels and defined them as follows:

\begin{itemize}[leftmargin=1.3cm]

    \item[\textbf{Wrong}] The document has nothing to do with the query, and does not help in any way to answer it.
  
    \item[\textbf{Topic}] The document talks about the general area or topic of a query, might provide some background info, but ultimately does not answer it.

    \item[\textbf{Partial}] The document contains a partial answer, but you think that there should be more to it.

    \item[\textbf{Perfect}] The document contains a full answer: easy to understand and it directly answers the question in full.

\end{itemize}

To make it easier for our non-expert annotators to distinguish the classes, we displayed them with strong visual cues as background colors of the voting buttons, as shown in Figure \ref{fig:labels}. Our annotation guidelines stated: ``For Partial and Perfect grades, you need to select the text spans that are in fact the relevant text parts to the questions.'' We purposefully kept the definition of ``relevant'' ambiguous to let the annotators decide for themselves what they deem relevant and to receive realistic relevance labels for non-expert web search engine users. Furthermore, we emphasised in our guidelines that if an answer requires domain specific knowledge and the connection to the query is not clear in the snippet, then that snippet is not relevant. Finally, we concluded our annotation guidelines with one example per relevance class.

\subsection{Annotation Campaign}

We conducted our annotation campaign among 87 computer science students. We set a target of 500 annotations per student and incentivized our students with bonus points if they continued after reaching their target. On average, each student annotated \numprint{890} samples.\footnote{We have to note that the annotation campaign was conducted during a COVID-19 lockdown, which might have helped us gather more annotations than we would have in a normal situation.}

Previous work highlighted the difficulty of working with student annotators \cite{palotti2016assessors}, therefore we took a number of steps to monitor the quality of the annotations: \textbf{1)} Besides a 3-way majority voting for every pair, we also let every student annotate the same set of 10 pairs spread throughout their workload; \textbf{2)} We monitored the time per annotation, to detect ``cheating'' by students who just immediately select any class; \textbf{3)} We asked the students for feedback twice (after the first 20 annotations and after they completed their target). 

Our campaign occurred over the course of 19 days as an exercise with a clear deadline for our students. It is generally believed that academics and students alike are especially motivated closer to a deadline. To better understand this phenomenon and to provide quality control, we plot the cumulative and running annotation trajectories, as well as the daily-average time per label over the course of our campaign in Figure \ref{fig:campaign-monitoring}. In the top plot, we see that some students finished their target earlier, however, most students finished close to the first deadline. In the second plot, we can clearly observe the deadline motivation phenomenon, with increased activity starting 5 days before the deadline. Even though we observe this increased activity, the average time per annotation only decreases gradually by a few seconds. The \textit{Partial} and \textit{Perfect} times are naturally higher than the other two classes, as the annotators were required to select text in addition to choosing a label.

Overall our students gave us very positive feedback for the question: \textit{How did you like to work with FiRA so far?} 38\% said \textit{Very good}, 45\% \textit{Good}, 12\% \textit{Decent}, and only 5 \% selected \textit{Don't like} as the rating feedback.\footnote{The feedback was not anonymous, so that we could follow up on the written feedback.} We see this as a positive confirmation of our task design. Many students also gave written feedback, which helps us tremendously to understand the limitations of our dataset. The main problem among annotators was the distinction between the \textit{Wrong} and \textit{Topic} labels. We believe that this problem arose from the fact that we only annotate documents that have at least somewhere in them a relevant part, making it more likely that the unrelated parts stay on topic. Therefore, when using the annotations, we should assume both classes to be non-relevant. Interestingly, if we consider the average time per annotation from Figure \ref{fig:campaign-monitoring} we observe that the \textit{Topic} label took a few second longer to classify than \textit{Wrong}, which leads us to believe that the two classes, while challenging and noisy to distinguish in single examples, make sense overall.

\begin{table}[t]
    \centering
    \caption{FiRA dataset statistics. The \# of labels is after majority voting.}
    \label{tab:fira_stats}
    \vspace{-0.3cm}
    \begin{tabular}{lr}
       
       \toprule
       \textbf{\# of queries} &  \numprint{43}    \\
       \textbf{\# of documents in collection} &  \numprint{3213835}    \\

       \textbf{\# of FiRA annotated query-document pairs} &  \numprint{1990}    \\
       \textbf{+ TREC-DL \# of query-document pairs} &  \numprint{16258}    \\
       \midrule
       \textbf{\# of judged document snippets}    &  \numprint{24197} \\
       \textbf{\# of total judgements}   &  \numprint{78340}  \\
       \midrule
       \textbf{\# of 0: Wrong}  &  \numprint{13565} (\numprint[\%]{56}) \\
       \textbf{\# of 1: Topic}  &  \numprint{6201} (\numprint[\%]{26}) \\
       \textbf{\# of 2: Partial}  &  \numprint{3145} (\numprint[\%]{13}) \\
       \textbf{\# of 3: Perfect}  &  \numprint{1286} (\numprint[\%]{5}) \\
       \midrule
       \textbf{\# of annotators} &  87  \\ 
       \textbf{avg. \# per annotator} &  890  \\ 

        \bottomrule
    \vspace{-.6cm}
    \end{tabular}
\end{table}

\subsection{FiRA Dataset Processing}

After the annotation campaign we post-processed the raw annotations to form the final published FiRA dataset. In Table \ref{tab:fira_stats}, we give an overview of the number of judged pairs, total annotations, and label distribution of the FiRA dataset. The combination of TREC-DL and FiRA creates a very densely annotated retrieval and QA-snippet dataset, which can be used by any approach investigating fine-grained relevance aspects of retrieval and QA.

We transform the raw annotations into standard-format qrels for retrieval, as well as a specialized but simple format for indicating character-level selection of relevant regions. We purposefully stay at a character-level to be independent of tokenization methods, however we deliver simple helper scripts to easily map our character ranges to tokens of any tokenization method.

\begin{figure}
    \centering
    \includegraphics[width=0.45\textwidth,trim=0 0 0 3cm ,clip]{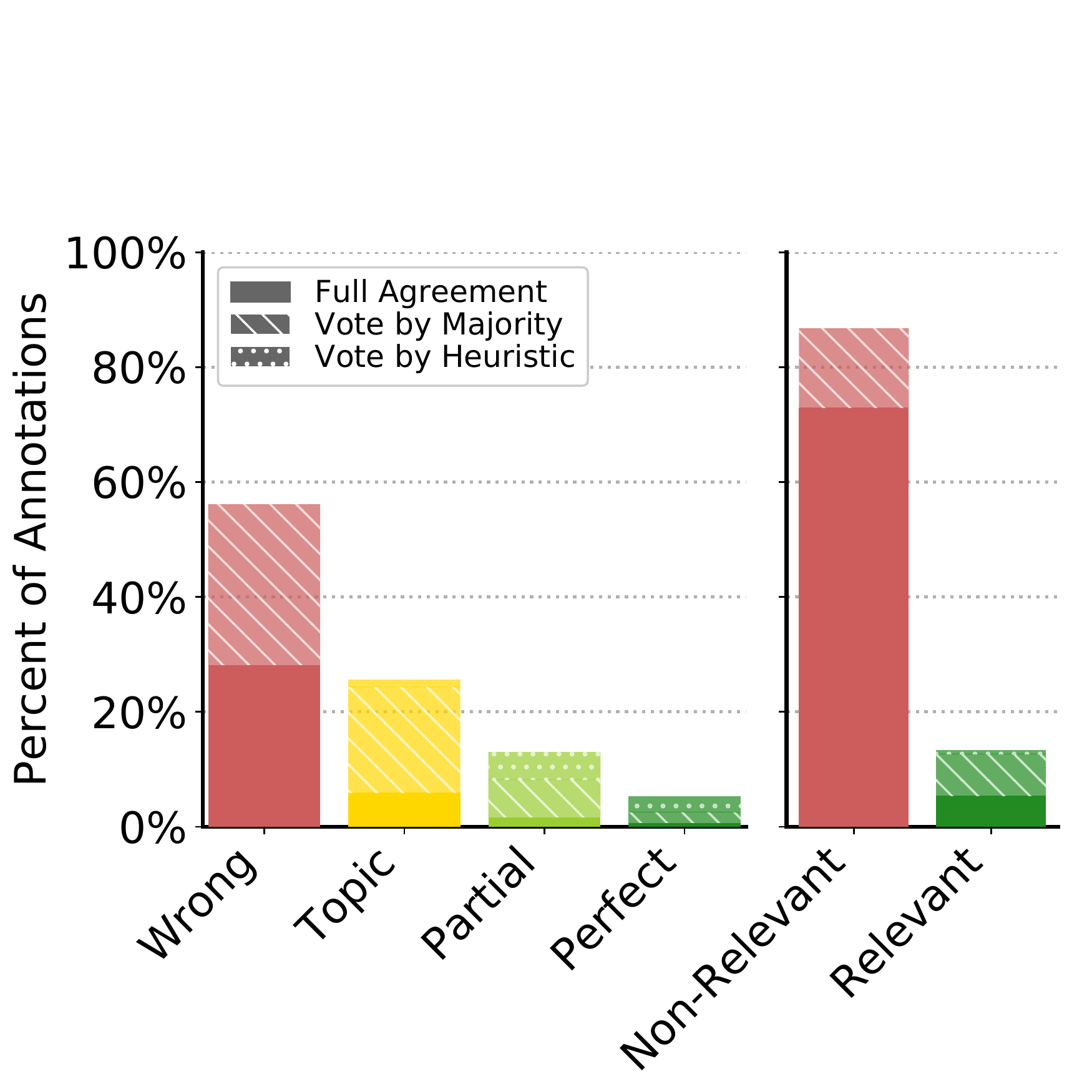}
    \vspace{-0.5cm}
    \caption{FiRA's distribution of labels, and ratio of heuristic \& majority voting.}
    \label{fig:label_distribution}
\end{figure}

To distill our raw annotations to voted judgements per query-document snippet, we relied on the following procedure: If a full agreement between all annotators or a majority for a single class exists, we take the majority class. If there exists a split between two or more labels, we employ the heuristic to take the highest order of relevance class.\footnote{We take the descending order of: Perfect, Partial, Topic, Wrong.} We apply this heuristic with the assumption that if an annotator took a higher class, they spent more time on the decision on average (as seen in Figure \ref{fig:campaign-monitoring}) and the two relevant classes require a selection of text, which reduces the risk of false positives. 

In Figure \ref{fig:label_distribution}, we show the distribution of classes in the dataset as well as the distribution of the judgement voting type per class. We rarely see a full agreement of all judges. This shows that our approach of prioritizing 3-way majority voting coverage over more pairs to annotate was warranted.

We have to emphasize that we only annotated documents judged to be overall relevant by TREC. We observe that even though some part of a document is relevant, most parts are not relevant to the query. We investigate this observation further in Section \ref{sec:pos_bias}.

\section{Agreement \& Subjectivity}
\label{sec:agreement}

When we treat retrieval collections solely as a given universal ground truth to measure retrieval metrics, we are prone to overlook the inherent subjectivity of their assignment and the annotators creating the dataset. As part of FiRA, we selected 10 pairs to be graded by all annotators, evenly distributed throughout the task. We selected the 10 query-document snippet pairs to contain (in our opinion) 3 \textit{Perfect}, 3 \textit{Partial}, 2 \textit{Topic} and 2 \textit{Wrong} instances. We conducted this experiment to obtain evidence towards answering RQ1: \textit{How much subjectivity do our fine-grained relevance annotations exhibit?}

We analyze the subjectivity based on the inter-annotator agreement (IAA) between the student annotators. We compute the agreement via Cohen's Kappa, a standard metric to compare two sets of annotations. As data basis, we consider the 10 pairs that were annotated by all students. Based on these 10 pairs, we compute an aggregated annotation by majority vote; then, we compare each student's annotation with the aggregated one.

\begin{figure}
    \centering
    \includegraphics[width=0.45\textwidth]{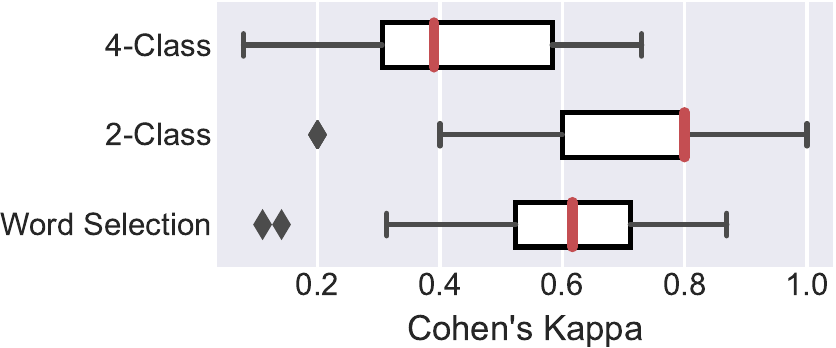}
    \vspace{-0.1cm}
    \caption{Cohen's Kappa agreement between individual students and the aggregated annotations via majority voting.}
    \label{fig:kappa}
    \vspace{-0.2cm}
\end{figure}

The Kappa scores for 2-classes,\footnote{We combine \textit{Wrong} and \textit{Topic}, as well as \textit{Partial} and \textit{Perfect}} 4-classes, and word-based relevance assignment are presented in Figure~\ref{fig:kappa}. For the 2-class relevance assignment and the labeling of the relevant word phrases, substantial Kappa agreements are reached, ranging from 0.5 to 0.8. For the more difficult task of differentiating between all four relevance classes, a mediocre agreement ranging from 0.3 to 0.6 is measured. Our results align with~\citet{alonso2012using} who report similar Kappa agreements for the relevance labeling of TREC collections when employing non-expert annotators.

The inter-annotator agreement is a metric for the quality in data annotation project. A low agreement indicates that the annotators failed to agree on common judgements. However, for the task of relevance feedback assignment for IR, also the subjectivity of individual workers is a source of disagreement \cite{borlund2003concept}: For example, one annotator might find a given text phrase relevant whereas a second annotator does not find that the exact same text phrase sufficiently answers the query. Therefore, for relevance judgement tasks, a certain level of disagreement can be expected since the task subjectivity needs to be taken into account as well.

\begin{figure}[!t]
    \noindent
    {\color[RGB]{76, 76, 90}Query} \textbf{what is physical description of spruce} \newline
    \line(1,0){242}

    \vspace{0.2cm}
    \begin{minipage}{.47\textwidth}
    \textbf{Perfect: 44}, Partial: 31, Topic: 1, Wrong: 3\newline
    \end{minipage}
    
    \begin{minipage}{.47\textwidth}
    \footnotesize
    \cb{18}{The} \cb{18}{trees} \cb{18}{have} \cb{18}{a} \cb{18}{number} \cb{18}{of} \cb{19}{} \cb{21}{key} \cb{21}{characteristics} \cb{18}{that} \cb{18}{help} \cb{18}{them} \cb{18}{stand} \cb{18}{out} \cb{18}{from} \cb{18}{their} \cb{18}{coniferous} \cb{18}{cousins:} \cb{22}{} \cb{59}{Leaves:} \cb{59}{} \cb{100}{Spruce} \cb{99}{trees} \cb{99}{feature} \cb{99}{stiff} \cb{99}{needles} \cb{99}{which} \cb{99}{range} \cb{99}{in} 
    \cb{99}{color} \cb{99}{from} \cb{99}{silvery-green} \cb{99}{to} \cb{99}{blue-green} \cb{96}{depending} \cb{96}{on} \cb{96}{the} \cb{96}{type} 
    \cb{96}{of} \cb{96}{specimen} \cb{82}{.} \cb{81}{} \cb{89}{The} \cb{89}{} \cb{92}{needles} \cb{92}{often} \cb{92}{} \cb{93}{curve} \cb{93}{inward} \cb{93}{} \cb{92}{and} \cb{92}{} \cb{93}{measure} \cb{93}{about} \cb{93}{three} \cb{93}{quarters} \cb{93}{of} \cb{93}{an} \cb{93}{inch} \cb{93}{long} \cb{62}{.} \cb{55}{} \cb{71}{Bark:} \cb{73}{} \cb{82}{The} \cb{82}{} \cb{84}{grayish-brown} \cb{84}{bark} \cb{84}{sports} \cb{84}{a} \cb{84}{moderate} \cb{84}{thickness} \cb{63}{.} \cb{53}{} \cb{58}{It} \cb{58}{forms} \cb{58}{furrows} \cb{48}{,} \cb{42}{} \cb{47}{ridges} \cb{47}{and} \cb{47}{scales} \cb{47}{as} \cb{47}{the} \cb{47}{tree} \cb{47}{matures} \cb{40}{.} \cb{37}{} \cb{63}{Fruit:} \cb{63}{} \cb{74}{Light} \cb{74}{brown} \cb{73}{,} \cb{73}{} \cb{74}{slender} \cb{74}{cones} \cb{73}{} \cb{74}{with} \cb{74}{diamond} \cb{73}{} \cb{74}{shaped} \cb{74}{scales} \cb{71}{} \cb{70}{contain} \cb{70}{seeds} \cb{68}{} \cb{66}{which} \cb{66}{are} \cb{66}{transported} \cb{66}{by} \cb{66}{the} \cb{66}{wind} \cb{47}{.} \cb{45}{} \cb{47}{The} \cb{47}{} \cb{48}{cones} \cb{48}{typically} \cb{48}{fall} \cb{48}{from} \cb{48}{the} \cb{48}{tree} \cb{48}{shortly} \cb{48}{after} \cb{48}{the} \cb{48}{seeds} \cb{48}{are} \cb{48}{dispersed} \cb{32}{.} \cb{27}{} \cb{37}{Another} \cb{38}{} \cb{40}{distinguishing} \cb{40}{characteristic} \cb{40}{of} \cb{40}{} \cb{41}{the} \cb{41}{} \cb{42}{Spruce} \cb{42}{tree} \cb{42}{is} \cb{42}{its} \cb{42}{longevity} \cb{29}{.} \cb{26}{} \cb{30}{Some} \cb{33}{types} \cb{33}{can} \cb{33}{} \cb{34}{live} \cb{34}{up} \cb{34}{to} \cb{34}{800} \cb{34}{years} \cb{33}{thanks} \cb{33}{to} \cb{33}{their} \cb{33}{} \cb{34}{ability} \cb{34}{to} \cb{34}{withstand} \cb{34}{extreme} 
    \cb{34}{weather} \cb{34}{conditions} \cb{19}{.}
      \newline
    \end{minipage}
    
    \line(1,0){242}
    \vspace{0.2cm}
    \begin{minipage}{.47\textwidth}
    Perfect: 19, \textbf{Partial: 53}, Topic: 4, Wrong: 1\newline
    \end{minipage}\vspace{-0.1cm}
    \begin{minipage}{.47\textwidth}
    \footnotesize
    \cb{3}{Types} \cb{3}{of} \cb{3}{Spruce} \cb{3}{Trees} \cb{2}{"Home} \cb{2}{»Trees} \cb{2}{Types} \cb{2}{of} \cb{2}{Spruce} \cb{2}{Trees} \cb{2}{By} \cb{2}{John} \cb{2}{Lindell;} \cb{2}{Updated} \cb{2}{September} \cb{2}{21,} \cb{2}{} \cb{6}{2017The} \cb{6}{} \cb{9}{spruce} \cb{9}{trees} \cb{9}{of} \cb{9}{North} \cb{9}{America} \cb{9}{include} \cb{9}{seven} \cb{9}{different} \cb{9}{species,} \cb{9}{all} \cb{9}{of} \cb{9}{them} \cb{9}{growing} \cb{9}{north} \cb{9}{of} \cb{9}{Mexico,} \cb{9}{mostly} \cb{9}{in} \cb{9}{cool} \cb{9}{climate} \cb{9}{locations.} \cb{9}{} \cb{14}{Spruce} \cb{14}{trees} \cb{14}{are} \cb{14}{important} \cb{14}{producers} \cb{14}{of} \cb{14}{lumber} \cb{14}{and} \cb{14}{are} \cb{14}{also} \cb{14}{useful} \cb{14}{in} \cb{14}{an} \cb{14}{ornamental} \cb{14}{capacity.} \cb{23}{} \cb{91}{The} \cb{91}{} \cb{95}{spruces} \cb{95}{all} \cb{95}{have} \cb{98}{} \cb{100}{evergreen} \cb{100}{needles} \cb{94}{,} \cb{94}{most} \cb{94}{are} \cb{92}{} \cb{95}{tall} \cb{55}{} \cb{60}{and} \cb{60}{} \cb{63}{they} \cb{65}{have} \cb{65}{a} \cb{65}{} \cb{66}{conical} \cb{66}{shape} \cb{26}{.} \cb{18}{} \cb{8}{White} \cb{8}{Spruce} \cb{8}{} \cb{11}{White} \cb{11}{spruce’s} \cb{11}{range} \cb{11}{is} \cb{11}{“transcontinental} \cb{9}{,”} \cb{8}{according} \cb{8}{to} \cb{8}{the} \cb{8}{Nearctica} \cb{8}{website,} \cb{8}{} \cb{9}{as} \cb{9}{the} \cb{9}{tree} \cb{9}{grows} \cb{9}{from} \cb{9}{Labrador} \cb{9}{to} \cb{9}{Alaska,} \cb{9}{covering} \cb{9}{most} \cb{9}{of} \cb{9}{Canada} \cb{9}{and} \cb{9}{many} \cb{9}{of} \cb{9}{the} \cb{9}{northernmost} \cb{9}{states.} \cb{17}{} \cb{92}{White} \cb{92}{} \cb{94}{spruce} \cb{94}{grows} \cb{94}{to} \cb{95}{150} \cb{95}{feet} \cb{95}{tall} \cb{94}{} \cb{95}{and} \cb{95}{} \cb{97}{features} \cb{97}{blue-green} \cb{97}{needles} \cb{60}{.} \cb{49}{} \cb{58}{The} \cb{58}{odor} \cb{58}{that} \cb{58}{crushed} \cb{58}{needles} \cb{57}{produce} \cb{57}{gives} \cb{57}{the} \cb{57}{tree} \cb{57}{the} \cb{57}{nickname} \cb{57}{of} \cb{57}{skunk} \cb{57}{spruce} \cb{25}{.}      
    \newline
    \end{minipage}
    \vspace{-0.2cm}
    \caption{Comparison of two completely judged document snippets. The background heatmap for each word displays the number of times this word was part of a relevant region selection. The darker the green the more often annotators selected this word as being relevant.}
    \label{fig:interpretability_2col}
    \vspace{-0.4cm}
\end{figure}

As a case based example for subjectivity, we present in Figure \ref{fig:interpretability_2col} the distribution of fine-grained word-level annotations on two document snippets for the same query. We would classify the snippet on top as \textit{Perfect} and the lower as \textit{Partial}. The majority of our annotators also selected the respective labels, although we see a strong overlap between the two relevant classes. We can observe that in some instances, such as the second snippet, it is very easy to select the relevant regions, and most annotators agree on two sentences, whereas in the first snippet we see a greater variability with the second and third sentence at its core. In our opinion, relevance is context dependent, especially in the second example of Figure \ref{fig:interpretability_2col}, as the selected answers are correct and fulfill the information need: if you seek a short and concise answer it would be \textit{Perfect}, however if someone seeks an exhaustive description of spruce trees it would only be \textit{Partial}.

\begin{figure*}
    \centering
    \includegraphics[width=0.93\textwidth]{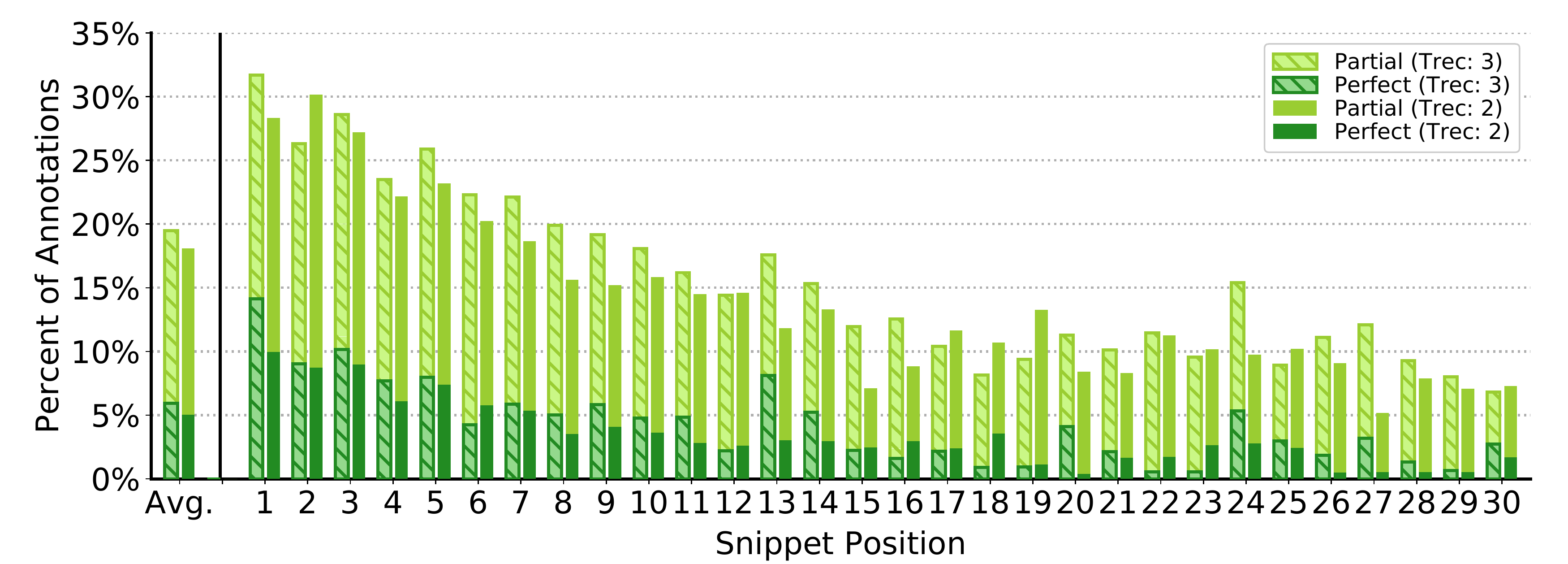}
    \vspace{-0.4cm}
    \caption{Relative amount of relevant annotations per position of the snippets in the document, for both TREC labels.}
    \label{fig:doc-pos-bias}
    \vspace{-0.4cm}
\end{figure*}

\section{Position Bias Study}
\label{sec:pos_bias}

In this study, we look at two different position bias scenarios. First, a bias of snippet relevance based on the location in the document, which is determined by a combination of how humans write documents, and the broad and general nature of the questions asked in the TREC-DL dataset. Our second focus is studying whether annotators pay insufficient attention and prematurely decide on the relevance of a snippet after reading only its beginning. To this end, we conducted an experiment during the annotation process.

\subsection{In-Document Position Importance}

Previous studies repeatedly highlighted the greater importance to an overall document relevance assessment of some passages, such as introductory paragraphs in news articles \cite{catena2019,wu2019investigating} and first-k term models in TREC-DL 2019 \cite{Hofstaetter2019_trec}. In our study, we aimed to manually confirm this observation in a web search context and answer our RQ2: \textit{Does there exist a position-based relevance bias in long documents?} To prevent an inherent bias against later positions, we showed our annotators snippets of documents in random order. With this approach, we are confident that our annotation results reflect true relevance differences and not user-interface based annotator bias.

Our results are detailed in Figure \ref{fig:doc-pos-bias}, where we show the proportion of relevant annotations per snippet location for documents split by TREC classes 2 (Highly Relevant) and 3 (Perfect). Our main observation here is a confirmation of these earlier studies that highlight the importance of earlier passages in a document. However, we also see that this distribution of relevance across longer documents is not very strongly skewed and a sizeable amount of relevant information can be found in later parts. This shows that for high-recall targets, also inside documents, it is vital to operate on the full documents.  

Our next observation based on Figure \ref{fig:doc-pos-bias} is that our fine-grained annotations only show a small difference for the TREC classes 2 (Highly Relevant) and 3 (Perfect) in terms of the likelihood of \textit{Perfect} and \textit{Partial} relevant snippets. The biggest difference occurs in the first snippet position, where TREC-3 has \numprint[\%]{4} more \textit{Perfect} annotations. This result can be the output of different annotation guidelines and setups, as we broke down and simplified the task, or just shows --- as discussed in Section \ref{sec:agreement} --- that we have to accept noise in relevance annotations due to human subjectivity. This result suggests dampening the gain values of relevance labels used in nDCG to be less different than 2 and 3, as we cannot observe a strong difference between the two relevant labels.

\begin{figure}
    \centering
    \includegraphics[width=0.48\textwidth]{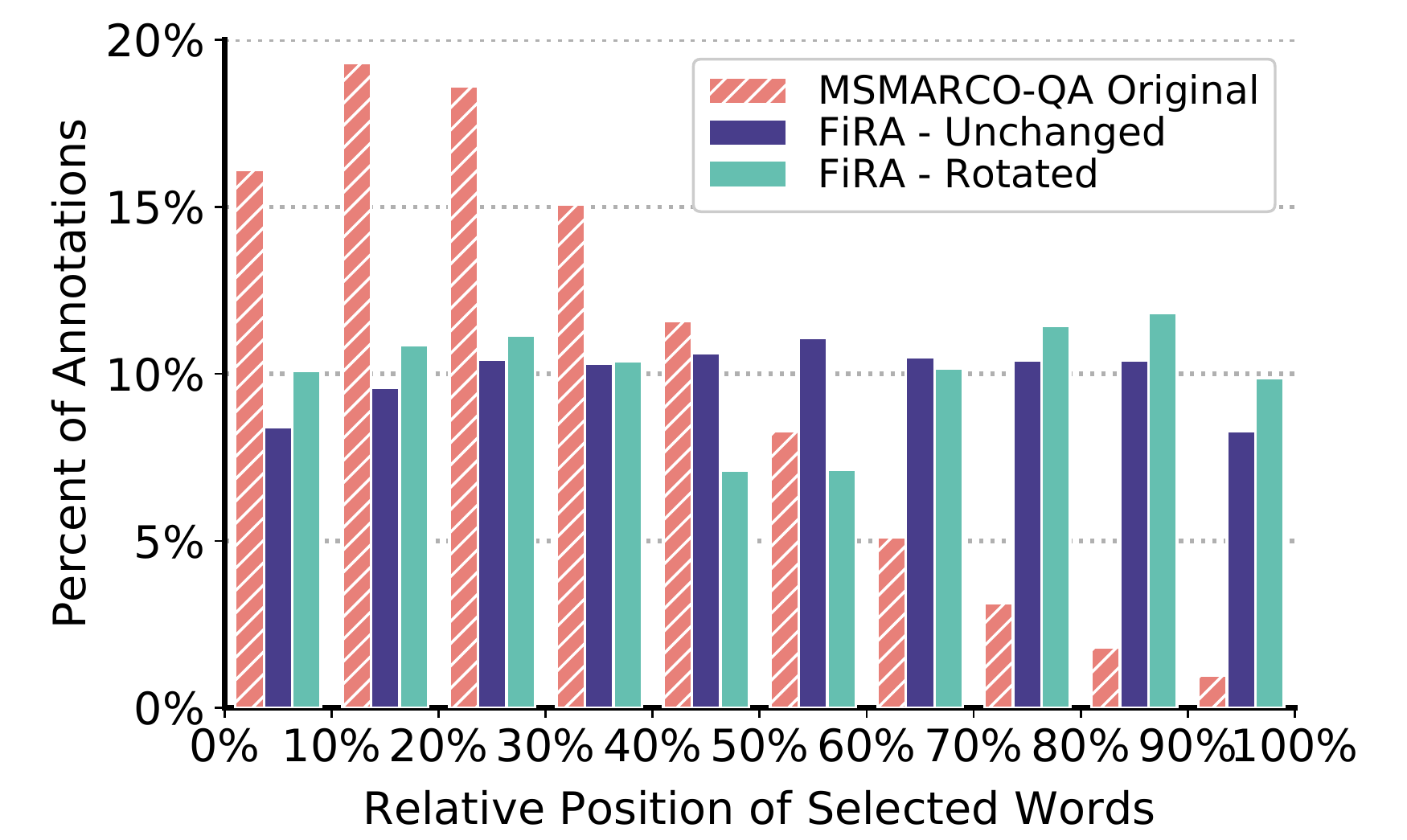}
    \vspace{-0.5cm}
    \caption{Relative distribution of selected words in passages of MSMARCO-QA and FiRA.}
    \label{fig:annotator-bias}
    \vspace{-0.5cm}
\end{figure}

\begin{figure*}
    \centering
    \includegraphics[width=0.95\textwidth]{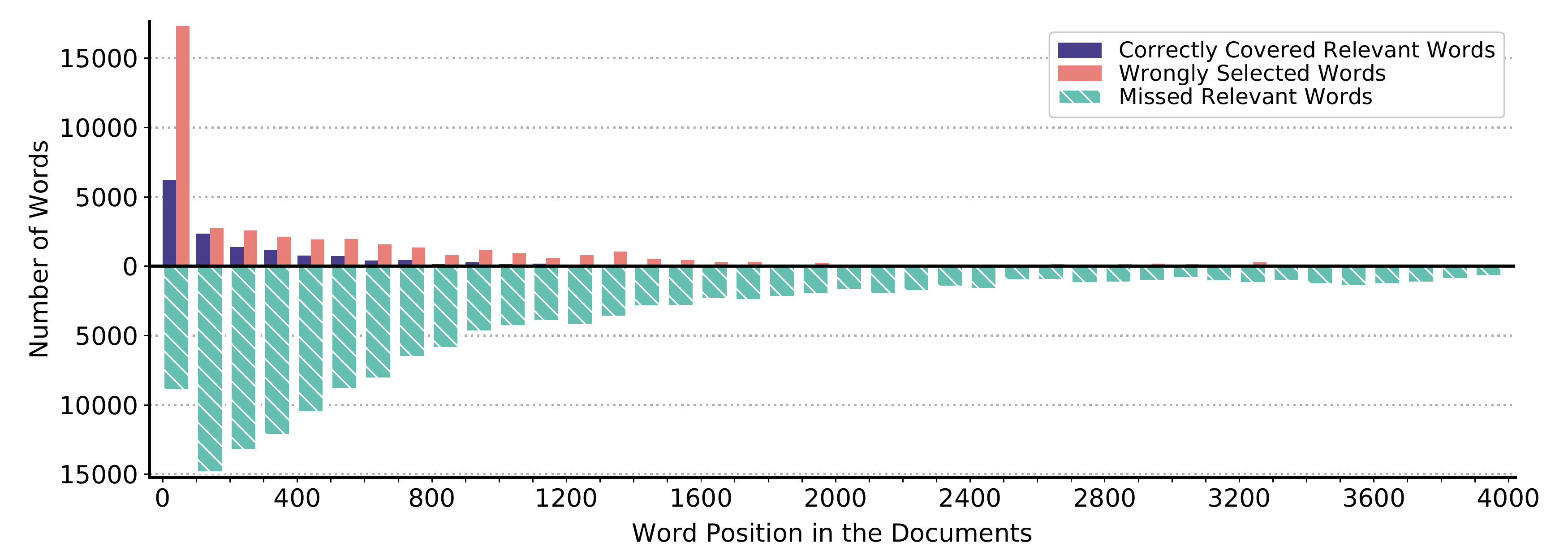}
    \caption{Analysis of TKL's relevant positions of retrieved relevant documents using FiRA. On top are the correctly (true-positive) and wrongly (false-positive) selected spans, whereas on the bottom is the ``iceberg'' of missed relevant spans selected in FiRA, but missed by TKL.}
    \label{fig:tkl-position-analysis}
    \vspace{-0.1cm}
\end{figure*}

\subsection{In-Snippet Annotator Bias}
\label{sec:pos_snip_bias}
Before starting our annotation campaign, we discovered an imbalance in the distribution of selected answer words\footnote{MSMARCO-QA does not contain exact answer spans, therefore we matched answer words to words in their respective training passages (matched: \numprint[\%]{32} or \numprint{176013} answers).} in the MSMARCO-QA dataset, which is the initial source of queries and websites later used in TREC-DL and FiRA.\footnote{The query passage pairs differ between MSMARCO-QA and FiRA, but the query distribution and domain remain the same} This imbalance is visible in Figure \ref{fig:annotator-bias}, where the MSMARCO-QA answers clearly favor earlier words in the passages. Our assumption is that this strong bias favoring earlier positions comes from annotators prematurely deciding that a passage is not relevant if the answer is not contained in the beginning. A major difference between MSMARCO-QA and FiRA is the annotation workflow: MSMARCO displayed a ranked list of results, whereas we displayed one document snippet at a time. 

To study RQ3 (\textit{Are annotators biased towards the beginning of a snippet?}) we rotated snippets half of the time before showing them to annotators. The rotation was conducted by cutting a snippet into two halves by the word count and concatenating the two halves in reverse order.
Then, after a judgement, we mapped the selection back to its original position in the document. Our results for the distribution of selected relevant words are shown in Figure \ref{fig:annotator-bias}.

Using our FiRA approach, in both the unchanged and rotated case we do not observe an imbalance of selected relevant words towards the beginning. In the unchanged passages, we see that the first and last \numprint[\%]{10} (roughly the first and last sentence) received fewer annotations. In the rotated case, we observe this dip in selected passages around the cutting point. Those words moved to the beginning and end of the user interface. 

Overall the average percentage of words in a snippet selected by an annotator is \numprint[\%]{31} with a standard deviation of \numprint[\%]{24} and a mean number of \numprint{43} selected words in total. This shows that our even position distribution does not come from annotators selecting every word.

Our annotators were not biased towards the beginning of passages, and in fact rotating the passages is not needed. We disproved our hypothesis, however we made another interesting discovery worth our attention in future work: Our annotators, using the FiRA tool, were less likely to select words from the first and last sentence of a snippet. We hypothesise that this is due to missing context information, cut off by the snippet creation. For future annotation campaigns, we plan to incorporate context sentences, connecting the current passage to the preceding and following ones, to further improve the uniform balance of selected answers. 

\section{Neural Document Ranking}

To showcase the usefulness of our FiRA dataset, we now turn to the analysis of neural re-ranking models. We focus in particular on the recently introduced TKL document ranking model \cite{Hofstaetter2020_sigir}, as it offers solid interpretation capabilities, namely it outputs detected relevant regions inside a long document.

The TKL model scores a query-document pair by first contextualizing the two sequences independently with Transformers, where the document sequence is split into overlapping chunks to accommodate long input sequences. The TKL model then creates a cosine match-matrix between every query and document term representation. This match matrix is scanned with a sliding saturation function, to detect the most relevant regions of a document. In addition to its ranking score, TKL outputs the position of the three top regions of a document, which were used as a basis to calculate the score. Now with FiRA, we can analyze, on a word-level, if the model focused on relevant regions or if the model learned artefacts that differ from human judgement. Hence we can answer our RQ4: \textit{How many of FiRA's relevant spans does the TKL model use to compute the document score?}

In Figure \ref{fig:tkl-position-analysis}, we analyze the word-level relevance prediction of the TKL model on the FiRA dataset. We plot the number of words that have been correctly marked as relevant, the number of words wrongly selected as relevant by TKL, and the number of missed relevant words that have not been captured by TKL. With FiRA data we can now visualize the ``iceberg'' of TKL's missed relevant words. The TKL model learns a strong bias towards the beginning of a document, as a ranking signal, although FiRA shows that much  relevant information is available later in the documents, where TKL rarely selects a region.

\section{Related Work}

Our work relates to datasets for question answering and retrieval as well as the analysis of document structure and passage importance in IR. In addition, we also highlight studies that could benefit from the usage of FiRA.

\vspace{-0.2cm}
\paragraph{\textbf{Datasets}}

The TREC document task \cite{trec2019overview} builds on re-crawled data from the MSMARCO-QA collection \cite{msmarco16}. The MSMARCO-QA collection differs from our work in that only single passages have been binary-annotated and the answers are natural language text written by the annotators.

The 2007 edition of WebCLEF featured a snippet selection evaluation, where annotators where asked to judge returned snippets from the participating systems \cite{web_clef2007}. However, due to the low participation count, and non-exhaustive evaluation the collection has been shown to not be reusable for future systems \cite{overwijk2008evaluation}. In contrast, we conducted an exhaustive and model agnostic annotation campaign.

In the landscape of Question Answering datasets, the SQuAD \cite{rajpurkar2016squad} QA dataset, was annotated by finding questions to fit an answer present in a given  Wikipedia passage. The SearchQA \cite{dunn2017searchqa} dataset expanded snippets with Google search result snippets. For larger context QA and advanced reasoning, \citet{clark2018think} published ARC and \citet{mihaylov2018can} published the OpenBookQA. \citet{naturalquestions2019} used real Google queries and annotated answers selected from Wikipedia articles in the Natural Questions dataset. Those datasets are focused on different aspects of QA, without necessary relevance annotations for a broader set of documents. Therefore, the transformation to IR collections results in very sparse relevance labels. FiRA tries to bridge this gap by extending TREC-DL ranking annotations with selected snippets.

Fine-grained annotations of long documents have already been the focus of other NLP domains such information extraction \cite{jain2020scirex} and relation extraction \cite{yao-etal-2019-docred}. Fine-grained annotations are valuable in specialised domains, such as medical tagging \cite{roberts2015role} or detecting proposition types in argumentation \cite{jo2020lrec}.

\vspace{-0.2cm}
\paragraph{\textbf{Relevance Analysis}}

Our work studies relevance distributions across a document using our new annotations as well as the subjectivity of the relevance task. To better understand the inherent meaning of judging relevance, \citet{inel2018studying} studied topical relevance and various settings of the ranking task, including different document granularities. 
\citet{wu2019investigating} studied the influence of passage-level relevance on full-document relevance for Chinese news articles. They found the first passage to be a very strong indicator of overall relevance.
\citet{Hofstaetter2020_ecir} created the Neural IR-Explorer to help us better understand single word-interactions in the score aggregation of neural re-ranking models. 

\vspace{-0.2cm}
\paragraph{\textbf{Neural Models}}

We identified numerous works on neural models published in recent years that would benefit from an evaluation and analysis based on FiRA. A common approach to utilizing the large-scale pre-trained BERT model \cite{devlin2018bert} in document ranking is to apply BERT to passages \cite{wu2020leveraging}, overlapping windows \cite{idst2019trecdl}, or single sentences \cite{yilmaz2019}. In all these cases, BERT produces partial results that need to be aggregated externally to produce a final ranking score that could be compared with traditional full-document judgements. With FiRA, these models could be evaluated on their respective granularity before aggregating scores. Apart from BERT-based approaches, \citet{jiang2019semantic} proposed a semantic text matching for long-form documents model, and \citet{zheng2019rltm} created RLTM for long document ranking. With modifications, both could also be analyzed with FiRA.

In addition to retrieval focused models, the research community also proposed multi-task models for the QA-at-scale or sometimes referred to open-world-QA task, where systems need to retrieve candidates and select answers. Models such as the Retriever-Reader model from \citet{ni2018learning}, the multi-task Retrieve-and-Read models \cite{nishida2018retrieve,nishida2019answering}, or models based on phrase indices \cite{seo2019real}, are mainly evaluated on re-purposed QA-collections that lack dense retrieval judgements. FiRA presents a perfect fit for the evaluation of such models, as it can cover both retrieval and question answering aspects equally well.

\section{Conclusion}
With the release of FiRA, we provide the research community with a high-quality dataset that can be employed in a range of diverse settings. We showcased the usage of FiRA with a position bias and distribution study, as well as fine-grained analysis of a neural document ranking model. We look forward to seeing novel approaches by the research community, from both retrieval and question answering fields, to evaluate and analyse their models with FiRA. 

\vspace{-0.2cm}
\paragraph{\textbf{Acknowledgements}} We want to thank Patrick Kerschbaum for implementing our custom annotation tool as well as our students of the Advanced Information Retrieval course in the summer term of 2020 for annotating the data and being so patient and motivated in the process. 

%
%

\bibliographystyle{ACM-Reference-Format}
\bibliography{references}

\end{document}